\documentclass[american,aps,pra,reprint,superscriptaddress,twocolumn]{revtex4-1}
\usepackage[unicode=true,pdfusetitle, bookmarks=true,bookmarksnumbered=false,bookmarksopen=false, breaklinks=false,pdfborder={0 0 0},backref=false,colorlinks=false] {hyperref}
\hypersetup{ colorlinks,linkcolor=myurlcolor,citecolor=myurlcolor,urlcolor=myurlcolor}
\usepackage{braket,colortbl,cleveref,amsthm,amsmath,amssymb,txfonts}
\definecolor{myurlcolor}{rgb}{0,0,0.7}
\usepackage{graphics,graphicx}
\usepackage{color}
\theoremstyle{plain}

\def\bea{\begin{eqnarray}}
\def\eea{\end{eqnarray}}
\def\ba{\begin{array}}
\def\ea{\end{array}}

\def\ket{\rangle}
\def\bra{\langle}
\def\beq{\begin{equation}}
\def\eeq{\end{equation}}

\begin{document}
\title{Thermodynamic utility of Non-Markovianity from the perspective of resource interconversion}

\author{Samyadeb Bhattacharya} 
\email{sbh.phys@gmail.com}
\author{Bihalan Bhattacharya}
\author{A. S. Majumdar}
\affiliation{S. N. Bose National Centre for Basic Sciences, Block JD, Sector III, Salt Lake, Kolkata 700 098, India}

%\author{Arun Kumar Pati$^{3}$ \footnote{...}}
%\affiliation{$^{3}$Harish-Chandra Research Institute, Chhatnag Road, Jhunsi, Allahabad - 211019, India \\}

\begin{abstract}
\noindent We establish a connection between non-Markovianity and negative entropy production rate for various classes of quantum operations. We analyse several aspects of unital and thermal operations in connection with resource theories of purity and thermodynamics. We fully characterize Lindblad operators corresponding to unital operations. We also characterize the Lindblad dynamics for a large class of thermal operations. We next generalize the definition of the entropy production rate for the non-equilibrium case to connect it with the rate of change of free energy of the system, and establish complementary relations between non-Markovianity and maximum loss of free energy. We naturally conclude that non-Markovianity in terms of divisibility breaking is a necessary resource for the backflow of other resources like purity or free energy under the corresponding allowed operations.

\end{abstract}

\maketitle

\section{Introduction}

Coupling to noisy environments ushers in the process of decoherence in  quantum mechanical systems. As a consequence, the system may monotonically relax towards the thermal equilibrium, or more generally, to non-equilibrium steady states \citep{alicki,lindblad,gorini,breuer}. This one way flow of information from the system to the environment is the signature of divisible evolution which is a direct consequence of the Born-Markov approximation \citep{breuer}. Under the assumption of a large stationary environment and the limit of weak system-environment coupling, it can be shown that the
Born-Markov approximation leads to the complete positive (CP) divisibility of the dynamics \citep{rivas1,breuerN,alonso}. On the contrary, beyond the limit of weak coupling and large stationary bath, this approximation is not valid, and CP-divisibility may break down causing non-Markovian information backflow \citep{blp1,proman1,proman2,proman3,proman4,proman5,proman8,proman9,proman10}. 

The backflow of information entailed in non-Markovian dynamics can essentially be converted into certain other resources in various information theoretic and thermodynamic protocols. For instance, it has been shown that non-Markovian information backflow allows perfect teleportation with mixed states \citep{task1}, efficient entanglement distribution \citep{task2}, improvement of capacity for large quantum channels \citep{task3} and efficient work extraction from a quantum Otto cycle \citep{task4}. In the above examples  it has been shown that non-Markovianity can be converted into resources such as entanglement, coherent information and free energy. Other possibilities of resource inter-conversion have also been proposed, such as conversion of  non-Markovianity  into purity \citep{sam1} and coherence \citep{sam2,sam3}. Information theories can be viewed as examples of resource inter-conversion \citep{M2}, and the examples mentioned above show us the power of non-Markovianity
  as a resource in quantum information theory and thermodynamics. 

The natural framework for formulating quantum thermodynamics is that of open quantum systems. Recent 
studies have revealed that memory effects spawning from
strong system-environment correlations leading to information
back-flow, can prolong the lifetime of quantum traits \citep{task1,task2,task3,task4}. Hence,
non-Markovianity can be regarded  as a useful resource for performing quantum information processing.  In a very recent work,  a formal resource theory of non-Markovianity  has been developed \citep{sam4}, where the information backflow  is considered as the resource. The immediate question that arises from the construction of such a resource theory is how non-Markovianity could be converted into various other resources within this framework. In this paper, we focus on the thermodynamic power of non-Markovianity (NM).

The main motivation of the present work is to
scrutinize NM in backdrop of the resource theory of purity \citep{purity1,purity2} and thermodynamics \citep{thermo1,thermo2}. One of the primary goals of this paper is to classify the Lindblad generators for various thermodynamic operations.  We connect NM with various resource theories via the notion of entropy production rate (EPR). EPR as defined later, is a fundamental quantity in non-equilibrium thermodynamics, whose positivity gives rise to the local form of the second law of thermodyanamics for open quantum system. The positivity implies that the system is approaching towards equilibrium. In this work we establish in the backdrop of various resource theoretic frameworks, that the negative EPR caused by the NM backflow of information indicates the situation where the system is driven away from equilibrium. We thus establish that NM necessarily drives the system away from the equilibrium.

We further examine more general quantum operations beyond the thermal maps, to study the NM effect on entropy production, where time dependent drivings (environment induced or externally applied) are present. Since, time dependent driving can itself induce resource regeneration, the usual entropy production rate (defined as EPR in the manuscript) cannot distinguish the information backflow solely caused by NM in such situations. We therefore adopt a more general definition of entropy production rate \citep{latest1},  (GEPR), suitable for certain driven open systems  \citep{latest1,latest2,latest3}. In what follows we discuss three different scenarios, each of which is more general than the previous. We start with the resource theory of purity, where unital operations are the allowed operations. We next consider a more general framework of resource theory of thermodynamics, with thermal operations as the allowed operations. Finally,  we study the case of
external driving by considering a generalized version of EPR. We show that the negativity of the entropy production is necessary for backflow of resources in each of these three situations.

The plan of our paper is as follows: Section II begins with a formal discussion on the entropy production rate. In the subsections  A and B, we investigate the thermodynamic aspects non-Markovian backflow of information in context of unital and thermal operations, respectively. In section III 
we extend our study beyond thermal operations, connecting the change of free energy with the 
generalized entropy production rate in this case.  In section IV we  present an example of
a spin bath model to validate our findings. We conclude in section V with a summary of our results. 
 
\section{Entropy Production Rate}

 EPR is defined as the negative time derivative of the relative entropy between the instantaneous state and the thermal state \citep{sphon}:
\beq\label{1}
\sigma(t)=-\frac{d}{dt}S(\rho(t)||\tau_{\beta}),
\eeq
where $S(A||B)=Tr[A(\ln A-\ln B)]$ is the von-Neumann relative entropy and $\tau_{\beta}=e^{-\beta H}/\mathcal{Z}$ is the thermal state of the system at inverse temperature $\beta$, where $\mathcal{Z}=Tr[e^{-\beta H}]$ is the partition function. Under thermal operation, the thermal state is the only fixed point of the dynamics. It can be shown that $\frac{d}{dt}S(\rho(t)||\tau_{\beta})=Tr[\dot{\rho}(t)\ln\rho(t)]+\beta Tr[H\dot{\rho}(t)]$, giving rise to the relation
\beq\label{2}
\sigma(t)=\frac{d}{dt}S(t)-\beta \mathcal{J}=\frac{d}{dt}\left(S(t)-\beta\mathcal{W}(t)\right),
\eeq
where $S(t)=-Tr[\rho(t)\ln\rho(t)]$ is the von-Neumann entropy and $\mathcal{J}= Tr[H\dot{\rho}(t)]$ is the heat current. $\mathcal{W}(t)=Tr[H(\rho(t)-\tau_{\beta})]$ is the maximum work that can be extracted from the system by applying thermal operation. EPR can also be generalized for R\'{e}yni divergence as $\sigma^{\gamma}(t)=-\frac{d}{dt}S_{\gamma}(\rho(t)||\tau_{\beta})$, where
$S_{\gamma}(\rho(t)||\tau_{\beta})=\frac{1}{\gamma-1}\ln\left(Tr[\rho(t)^{\gamma}\tau_{\beta}^{1-\gamma}]\right)~(\gamma >0),$ is the R\'{e}yni relative entropy.
We show that under unital operation and thermal operation, $\sigma^{\gamma}(t)$ is positive under divisible CPTP evolution. 

\subsection{Non-Markovian backflow of information under unital operations}

We first consider the backdrop of resource theory of purity \citep{purity1,purity2}. A successful approach to the fundamental aspects of thermodynamics can be adopted, by considering purity as a resource. This can be done from different operational perspectives, depending on the set of of easily implementable free operations. One  convincing approach towards this end is to consider noisy or unital operations as the free operations. In the following theorem we analyse the characteristics of Lindblad operators \citep{lindblad} for unital operations.\\\\
\noindent\textbf{Theorem 1:} \textit{For all  unital dynamical maps having corresponding Lindblad generators, the Lindblad operators are normal. }

\proof: We first prove the sufficiency condition, {\it viz}., if the Lindbladians are normal then the dynamics is unital.
Let us consider a dynamical map: $\rho(t)=\Lambda(\rho(0))$ having a Lindbladian generator $\dot{\rho}(t)=\mathcal{L}(\rho(t))=\sum_{\alpha=1}^{n\leq d_S^2}\Gamma_{\alpha}(t)\left(A_{\alpha}\rho(t)A_{\alpha}^{\dagger}-\frac{1}{2}A_{\alpha}^{\dagger}A_{\alpha}\rho(t)-\frac{1}{2}\rho(t)A_{\alpha}^{\dagger}A_{\alpha}\right)$. To show the unitality of the map it is enough to show that $\mathcal{L}(\mathbb{I}_S)=0$, where $\mathbb{I}_S$ stands for identity matrix corresponding to the system dimension.
%Hence we have $\mathcal{L}(\mathbb{I_S})=0$, where $\mathbb{I}_S$ is the identity matrix for the dimension of the system $S$.
Putting $\rho(t)=\mathbb{I}_S$ in the Lindblad master equation we have,
 \[\mathcal{L}(\mathbb{I}_S)= \sum_{\alpha=1}^{n\leq d_S^2}\Gamma_{\alpha}(t)\left(A_{\alpha}A_{\alpha}^{\dagger}-\frac{1}{2}A_{\alpha}^{\dagger}A_{\alpha}-\frac{1}{2}A_{\alpha}^{\dagger}A_{\alpha}\right)\]
 
Using the property of normality $A_{\alpha}A_{\alpha}^{\dagger}=A_{\alpha}^{\dagger}A_{\alpha}~~\forall \alpha$, we have $\mathcal{L}(\mathbb{I}_S)=0$. \\
  
We now prove the necessity condition, {\it viz.}, if the dynamics is unital then the Lindblad operators are normal.
Let us now consider $A_{\alpha}=\sum_{ij\alpha}C^{\alpha}_{ij}|i\ket\bra j|,$ where $|i\ket,~|j\ket$ forms a complete set of orthonormal basis vectors.
Now,
\[
\begin{array}{ll}
\mathcal{L}(\mathbb{I}_S)=\sum_{\alpha }\Gamma_{\alpha}(t)(A_{\alpha} A_{\alpha}^{\dagger}-A_{\alpha}^{\dagger} A_{\alpha}),\\
=\sum_{ijkl\alpha}\Gamma_{\alpha}(t)\left(C^{\alpha}_{ij}|i\ket\bra j| (C^{\alpha}_{kl})^{\ast}|l\ket\bra k|-(C^{\alpha}_{il})^{\ast}|l\ket\bra k|C^{\alpha}_{ij}|i\ket\bra j|\right),
\\
=\sum_{ijk\alpha}\Gamma_{\alpha}(t) C^{\alpha}_{ij}(C^{\alpha}_{kj})^{\ast}|i\ket\bra k|-\sum_{ijl\alpha}\Gamma_{\alpha}(t) C^{\alpha}_{ij}(C^{\alpha}_{il})^{\ast}|l\ket\bra j|,
\\
=\sum_{ik}\left(\sum_{j\alpha}\Gamma_{\alpha}(t) C^{\alpha}_{ij}(C^{\alpha}_{kj})^{\ast}\right)|i\ket\bra k|-\sum_{lj}\left(\sum_{i\alpha}\Gamma_{\alpha}(t) C^{\alpha}_{ij}(C^{\alpha}_{il})^{\ast}\right)|l\ket\bra j|,
\\
=\sum_{ik}\Lambda_{ik}(t)|i\ket\bra k|-\sum_{lj}\Lambda_{jl}(t)|l\ket\bra j|,\\
=\sum_{mn}\Lambda_{mn}(t)|m\ket\bra n|-\sum_{mn}\Lambda_{nm}(t)|m\ket\bra n|,
\\
=\sum_{mn}(\Lambda_{mn}(t)-\Lambda_{nm}(t))|m\ket\bra n|,
\end{array}
\]
Therefore, $\mathcal{L}(\mathbb{I}_S)=0$ implies $\Lambda_{mn}=\Lambda_{nm}$ $\forall n,~m$.
%Let us consider $A_{\alpha}=A_{jk}=|j\ket\bra k|$. Therefore $\sum_{\alpha=1}^{n}\Gamma_{\alpha}(t)L_{\alpha}=\sum_{jk}\Gamma_{jk}(t)\left[|j\ket\bra j|-|k\ket\bra k|\right]=\sum_{jk}(\Gamma_{jk}(t)-\Gamma_{kj}(t))|j\ket\bra j|=0$. Therefore we have $\Gamma_{jk}(t)=\Gamma_{kj}(t),~~\forall j,k$.
 Consider $A_{jk}=|j\ket\bra k|$. The Lindblad evolution for the unital channel can be expressed as
\[
\begin{array}{ll}
\mathcal{L}(\rho(t))=\frac{1}{2}\sum_{j,k}\Lambda_{jk}(t)\left(A_{jk}\rho(t)A_{jk}^{\dagger}-\frac{1}{2}A_{jk}^{\dagger}A_{jk}\rho(t)-\frac{1}{2}\rho(t)A_{jk}^{\dagger}A_{jk}\right)\\
~~~~~+\frac{1}{2}\sum_{j,k}\Lambda_{kj}(t)\left(A_{jk}^{\dagger}\rho(t)A_{jk}-\frac{1}{2}A_{jk}A_{jk}^{\dagger}\rho(t)-\frac{1}{2}\rho(t)A_{jk}A_{jk}^{\dagger}\right),
\end{array}
\]
since $A_{jk}^{\dagger}=A_{kj}$. Using $\Lambda_{mn}=\Lambda_{nm}$ $\forall n,~m$, this equation can be modified to 
\[
\begin{array}{ll}
\mathcal{L}(\rho(t))=\frac{1}{2}\sum_{jk}\Lambda_{jk}(t)\left(H_{jk}\rho(t)H_{jk}^{\dagger}-\frac{1}{2}H_{jk}^{\dagger}H_{\alpha}\rho(t)-\frac{1}{2}\rho(t)H_{jk}^{\dagger}H_{jk}\right)\\
+\frac{1}{2}\sum_{jk}\Lambda_{jk}(t)\left(\bar{H}_{jk}\rho(t)\bar{H}_{jk}^{\dagger}-\frac{1}{2}\bar{H}_{jk}^{\dagger}\bar{H}_{jk}\rho(t)-\frac{1}{2}\rho(t)\bar{H}_{jk}^{\dagger}\bar{H}_{jk}\right),
\end{array}
\]
where $H_{jk}=\frac{A_{jk}+A_{jk}^{\dagger}}{\sqrt{2}}$ and $\bar{H}_{jk}=\frac{i(A_{jk}-A_{jk}^{\dagger})}{\sqrt{2}}$ are Hermitian operators which are normal. \qed

This theorem completely characterizes the Lindbladians for unital operations. Using the theorem, we prove the following corollary.\\
\noindent\textbf{Corollary 1:} \textit{For unital quantum dynamical processes, NM is necessary to drive the system away from equilibrium. }

\proof: A unital operation can always be represented by the dynamical map: 
\[\Lambda_{U}(\rho(0)) = Tr\left[V_{SB}\left(\rho(0)\otimes\frac{\mathbb{I_B}}{d_B}\right)V_{SB}^{\dagger}\right], \]
where $V_{SB}$ is a global unitary process acting over the total system-environment state, $\mathbb{I}_B$ is the identity matrix for the  environment $B$, and $d_B$ is the dimension of the environment. For the mentioned operation, $\mathbb{I}_S/d_S$ is the fixed point, which corresponds to the thermal state at infinite temperature ($\beta\rightarrow 0$). Here $d_S$ is the dimension of the system. For unital evolution, we use the identity 
$S(\rho||\frac{\mathbb{I}}{d_S})=\ln d_S-S(\rho)$,
to show that the entropy production rate can be defined as $\sigma_U(t)=\frac{d}{dt}S(t)$.\\

% Therefore, for unital evolution, the entropy production rate can be defined as $\sigma_U(t)=\frac{d}{dt}S(t)$.  Alternatively we can also use the identity $S(\rho||\frac{\mathbb{I}}{d_S})=\ln d_S-S(\rho)$, to prove the same.
  The result holds for generalized R\'{e}yni entropy also. In a previous work \citep{abe1}, it has been shown that, for the Lindblad type evolution of generalized R\'{e}yni entropy $S_{\gamma}(t)=\frac{1}{1-\gamma}\ln Tr[\rho^{\gamma}(t)],~(\gamma > 0)$, we have 
\[\frac{d}{dt}S_{\gamma}(t)=2\sum_{\alpha}\Gamma_{\alpha}^U(t)\chi_{\alpha}(t),\]
where \[\chi_{\alpha}(t)=\frac{\gamma}{1-\gamma}\frac{1}{Tr[\rho^{\gamma}(t)]}Tr\left(\rho^{\gamma-1}(t)A_{\alpha}\rho(t)A_{\alpha}^{\dagger}-\rho^{\gamma}(t)A_{\alpha}^{\dagger}A_{\alpha}\right).\]
It has been proved previously \citep{abe1} that \[\chi(t)>\bra[A_{\alpha}^{\dagger}, A_{\alpha}]\ket_{\gamma},\] where $\bra X\ket_{\gamma}=\frac{Tr[X\rho^{\gamma}(t)]}{Tr[\rho^{\gamma}(t)]}$. Since for unital evolutions, $A_{\alpha}$s are normal, we always have $\chi(t)>0$. Therefore the generalized EPR $\sigma^{\gamma}_{U}(t)\geq 0$, for all divisible evolutions ($\Gamma_{\alpha}^U(t)\geq 0~~\forall \alpha$). Therefore $\sigma^{\gamma}_{U}(t)$ can only be negative when divisibility of the dynamical process breaks down ($\Gamma_{\alpha}^U(t)\leq 0$). This proves that NM is necessary to drive the system away from equilibrium. \qed

Following \textbf{Corollary 1}, we present an important result connecting NM information backflow and resource theory of purity.\\\\
\textbf{Result 1:} The rate of change of purity under a unital dynamical process can be represented as
\beq\label{6}
\begin{array}{ll}
\frac{dP}{dt}=-\sum_{\alpha}\Gamma_{\alpha}^U(t)Q(A_{\alpha}),\\
\end{array}
\eeq
where $\Gamma_{\alpha}^U(t)$ and $A_{\alpha}$ are respectively the Lindblad coefficients and Lindblad  operators for unital evolution. The asymmetry of an operator with respect to a quantum state can be defined as $Q(O_i)=||[\rho,O_i]||_{HS}^2$, where $||.||_{HS}$ denotes the Hilbert-Schmidt norm.\\
\proof: The purity of a state is given as: $P(t)=Tr[\rho^2]$. Therefore we have \[\frac{d}{dt}P(t)=2Tr[\rho(t)\frac{d\rho(t)}{dt}]=2Tr[\rho(t)\mathcal{L}(\rho(t))]\]
Using the the property of normal operator: $A_{\alpha}^{\dagger}A_{\alpha}=A_{\alpha}A_{\alpha}^{\dagger}$ and the cyclic property of trace, we get
\[
\begin{array}{ll}
Q(A_{\alpha})=Tr\left[(\rho(t)A_{\alpha}-A_{\alpha}\rho(t))(A_{\alpha}^{\dagger}\rho(t)-\rho(t)A_{\alpha}^{\dagger})\right],\\
~~~~~~=-2Tr\left[\rho(t)\left(A_{\alpha}\rho(t)A_{\alpha}^{\dagger}-\frac{1}{2}\{\rho(t),A_{\alpha}^{\dagger}A_{\alpha}\}\right)\right].
\end{array}
\]
Thus, using the form of $\mathcal{L}_U(\rho(t))$, we get
\[\frac{d}{dt}P(t)=-\sum_{\alpha}\Gamma_{\alpha}^U(t)Q(A_{\alpha}).\] \qed

Therefore, from \textbf{Result 1}, it is evident that purity can only regenerated in the NM region ($\Gamma_{\alpha}^U(t)<0$) of the dynamics.  It shows that in the backdrop of resource theory of purity, NM can be converted into purity, via information backflow.\\

We illustrate Result 1 by the following example. Let us consider the qubit depolarising dynamics with the corresponding Lindblad master equation given by
\[ \dfrac{d\rho}{dt}=\sum_i\Gamma_i(t)(\sigma_i\rho\sigma_i-\rho),\]
where $\sigma_i$s $(i=x,y,z)$ are Pauli operators. Considering a non-Markovian model presented
in Ref. \citep{victor}, the Lindblad coefficients are taken to be 
$\Gamma_i(t)=\Gamma(t) = e^{-t}\cos(t)$, for all $i$. We plot the rate of change of purity and $\Gamma(t)$ with respect to time. We find that the rate of change of purity is positive, only when the coefficients $\Gamma_i(t)$s are negative, validating our  findings. 

\begin{figure}[htb]
	{\centerline{\includegraphics[width=5cm, height=4.5cm] {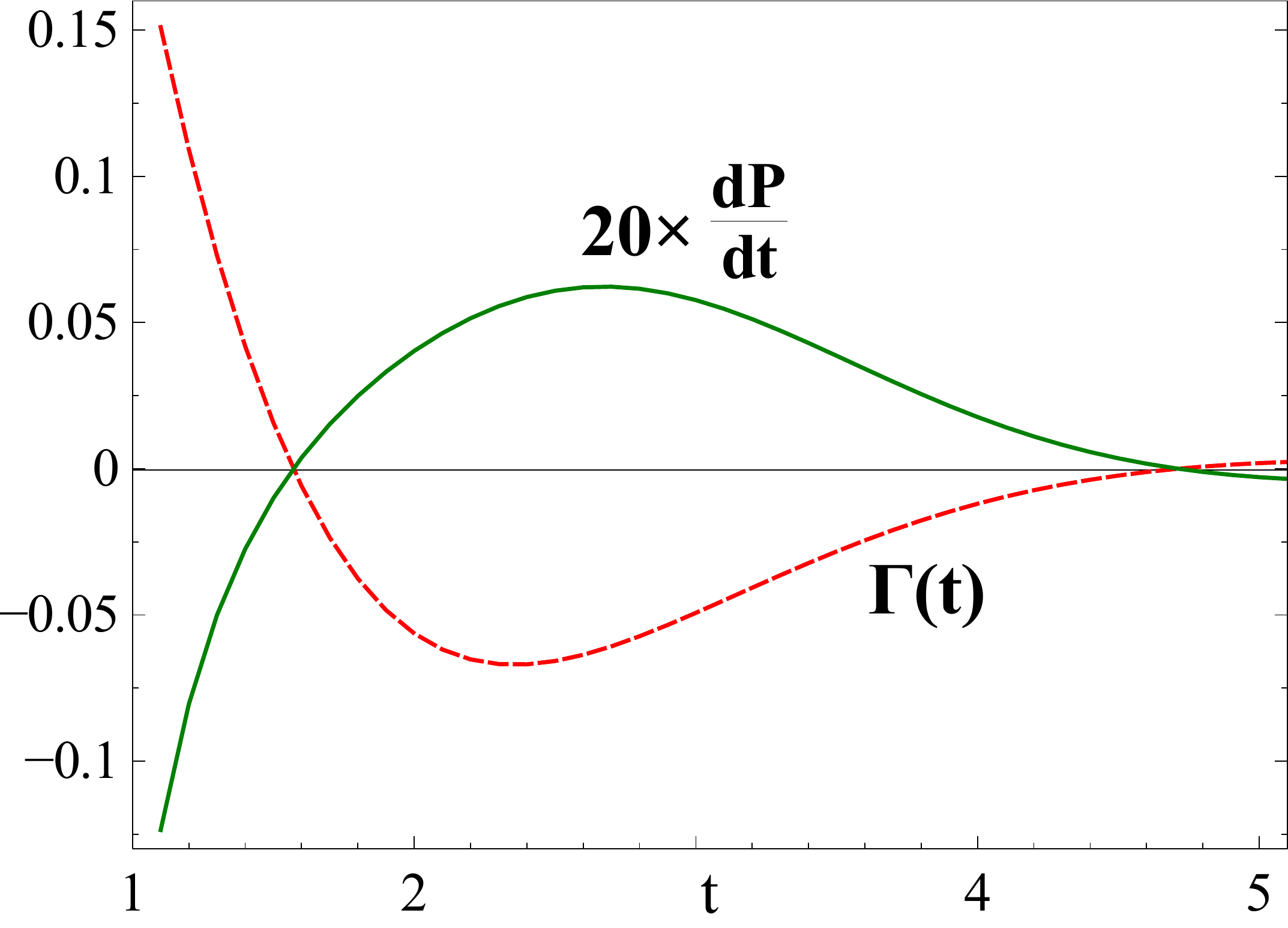}}}
	\caption{(Colour online)\\ In the Figure, we plot the time rate of change of purity $\frac{dP}{dt}$ (green line) and Lindblad coefficient $\Gamma(t)$ (red dashed line) with time. We observe that the rate of change of purity is positive, when $\Gamma(t)$ is negative; i.e. only in the non-Markovian situations.  }
	\label{fig1}
\end{figure}

\subsection{Non-Markovian backflow under thermal operations }

Let us now extend our study of classifying Lindblad dynamics for thermal operations. An operation $\Lambda_{th}$ is said to be thermal if the thermal state of the system $\tau_{\beta}$ is the fixed point of the operation and the free energy is a monotonically decreasing function under the same operation. Mathematically, a thermal operation is described as 
\beq\label{F1}
\Lambda_{th}(\rho(0)) = Tr\left[\mathcal{V}_{SB}\left(\rho(0)\otimes\mu_{\beta}\right)\mathcal{V}_{SB}^{\dagger}\right], 
\eeq
where $\mathcal{V}_{SB}$ is a energy preserving unitary operation acting on the system-environment composite Hilbert space and $\mu_{\beta}$ is the thermal state of the environment at a given temperature $\beta$.
The energy preservation condition invokes the following relation $\left[\mathcal{V}_{SB}, H_S+H_B\right]=0$, where $H_S$ and $H_B$ are the system and environment Hamiltonian respectively. Under these restrictions over the allowed operations, it can be shown that the thermal state of the system $\tau_{\beta}$ is the fixed point of the dynamics, i.e. $\Lambda_{th}(\tau_{\beta})=\tau_{\beta}$. Nevertheless, the Stinespring form \citep{alicki,breuer} of thermal operation given in Eq. \eqref{F1} is true for large class of unitary operators and it does not shed much light on the realistic constraints in experimental situations. Hence, our first goal in this subsection is the classification of the Lindblad generators corresponding to the thermal operations $\Lambda_{th}$.
In this context it is  important to note that full characterization of Lindblad generators corresponding to an arbitrary thermal operation is quite a difficult task since the class of energy preserving unitary operations is very large and there is no unique way to capture all of them. There have been attempts to bypass this problem by considering physically realizable dynamics, such as conditional thermal operations \citep{cto} or elementary thermal operations \citep{eto}. In this paper, we  identify a significant portion of thermal operations having clear experimental significance. In the following theorem, by proving a sufficient condition for a Lindblad operation to be thermal, we identify a particular subset of thermal operations.

\textbf{Theorem 2:} \textit{If the Lindblad operators are restricted to be of the form of rank 1 projector: $A_{ij}=|j\ket\bra i|$, with $\{|i\ket\}$ being the energy eigenbasis and $\Gamma_{ij}(t)$s  the corresponding Lindblad coefficients, then the condition for the operation to be a thermal operation is the detailed balance condition}: \[ \Gamma_{ji}(t)=\Gamma_{ij}(t)e^{-\beta(E_i-E_j)}\]

\proof Let us consider $\mathcal{L}_{th}$ to be the Lindblad generator corresponding to the thermal operation $\Lambda_{th}$. Gibbs preservation condition gives $\Lambda_{th}(\tau_{\beta})=\tau_{\beta}$. Therefore
$\dfrac{d\tau_{\beta}}{dt}=\mathcal{L}_{th}(\tau_{\beta})=0 $,

 yields $\mathcal{L}_{th}(\tau_{\beta})=0$.\\
Now, $\mathcal{L}_{th}(\rho)$= $\sum_{ijk} \Gamma_{ij}(t)(\vert j \ket\bra i \vert \rho \vert i \ket\bra j \vert - \dfrac{1}{2} \vert i \ket\bra i \vert \rho - \dfrac{1}{2} \rho \vert i \ket\bra i \vert)$.
Therefore, 
\[
\begin{array}{ll}
\Lambda_{th}(\tau_{\beta})=
\linebreak \sum_{ijk} \Gamma_{ij}(t) (\vert j \ket\bra i \vert e^{-\beta E_{k}} \vert k \ket\bra k \vert i\ket\bra j\vert -\dfrac{1}{2} \vert i \ket\bra i \vert k \ket\bra  k\vert e^{-\beta E_{k}}\\
-\dfrac{1}{2} e^{-\beta E_{k}} \vert k\ket\bra k \vert i \ket \bra i \vert) 
\end{array}
\]
 %=\sum_{ij}(e^{-\beta E_{i}} \vert j \ket\bra j \vert - e^{\beta E_{i}} \vert i \ket\bra i \vert  )
  \[
\begin{array}{ll}
 =\sum_{ij}( \Gamma_{ij}(t)e^{-\beta E_{i}} - \Gamma_{ji}(t) e^{-\beta E_{j}}) \vert j \ket\bra j \vert
\end{array}
 \]
Hence, $ \Lambda_{th}(\tau_{\beta}) = 0$ implies $\Gamma_{ji}(t) =e^{-\beta (E_{i}- E_{j})} \Gamma_{ij}(t)$ \qed\\

%i.e. $\Gamma_{ji} = \Gamma_{ij} e^{-\beta (E_{i}- E_{j})}$  

As a relevant example, consider a simple Markovian model, with a qubit is weakly coupled to a thermal bosonic environment. In absence of any external driving, the qubit eventually thermally equilibrates with the environment.
Under Born-Markov approximation, the master equation for this model is given by 
\beq\label{sec1fA}
\begin{array}{ll}
\dot{\rho}(\tilde{t})=\frac{i}{\hbar}[\rho(\tilde{t}),H_0]+\gamma(n+1)\left(\sigma_{-}\rho(\tilde{t})\sigma_{+}-\frac{1}{2}\{\sigma_{+}\sigma_{-},\rho(\tilde{t})\}\right)\\
~~~~~~~+\gamma n\left(\sigma_{+}\rho(\tilde{t})\sigma_{-}-\frac{1}{2}\{\sigma_{-}\sigma_{+},\rho(\tilde{t})\}\right),
\end{array}
\eeq
where $H_0=\hbar\Omega_0|1\ket\bra 1|$ is the Hamiltonian of the system, $\gamma$ is a constant parameter and $n=1/(\exp(\hbar\Omega_0/K\tilde{T}_m)-1)$
is the Planck number. Here $\sigma_{+}$ and $\sigma_{-}$ are respectively the raising and lowering operators of the two level system,
with $|1\rangle$ being the excited state of the same. This operation represents a thermal operation  on a two level system. The two Lindblad operators corresponding to the operation are rank one projectors and hence satisfy all the properties stated in \textbf{Theorem 2}.  It is strightforward to check, that the thermal state of the qubit corresponding to the bath temperature $\hat{T}_m$ and the system Hamiltonian $H_0$ is the only fixed point of this dynamics, proving the operation to be thermal. 

At this stage, it is interesting to compare our class of thermal operations having rank 1 projectors as Lindblad operators, with a physically implementable sub-class of thermal operations, namely elementary thermal operations \citep{eto}. It has been previously shown \citep{eto}, that elementary thermal operations are those which satisfy the following two criteria:, (i) the map involves only two energy levels of the system, and (ii) it satisfies the detailed balance condition.
It is clear from \textbf{Theorem 2} that if only two particular energy levels $(i,j)$ are involved in the Lindblad type evolution, the sub-class of thermal operations that we have considered is nothing but the class of elementary thermal operations. The consequence of this finding is important from experimental perspectives. Two-level population dynamics  can generally be realized by elementary thermal operations involving a single mode bosonic bath, and interestingly, the Jaynes-Cummings model can reproduce them to a satisfactory extent \citep{eto}. Therefore, it is evident that the class of thermal operations we consider in \textbf{Theorem 2} encompasses a considerable number of elementary thermal operations which are physically realizable in experimental situations. 

We now focus on  NM and its importance from the perspective of resource inter-conversion.   
A very relevant aspect of quantum information theory is the study of interconversion of different resources. Quantum non-Markovianity is one of the resources which  can  be studied from the perspective of quantum thermodynamics. Here it becomes important to observe how  different thermodynamic quantities respond under the presence of non-Markovianity. In order to do so, 
let us consider the role of NM in the resource theory of thermodynamics. Under thermal operations, the free energy $\mathcal{F}(t)=\bra H\ket-S(t)/\beta$ is a monotone and we prove that it obeys the following relation with EPR. 
\beq\label{3}
\beta\frac{d}{dt}\mathcal{F}(t)=-\sigma(t).
\eeq

The proof follows from the observation that, the free energy can be expressed as $\mathcal{F}(t)=\frac{1}{\beta}S(\rho(t)||\tau_{\beta})-\frac{1}{\beta}\ln\mathcal{Z}$. Therefore under thermal operation, we have $\beta\frac{d}{dt}\mathcal{F}(t)=\frac{d}{dt}S(\rho(t)||\tau_{\beta})=-\sigma(t).$ 

A more general definition of free energy \citep{thermo2} can be stated as $\mathcal{F}^{\gamma}(t)=\frac{1}{\beta}S^{\gamma}(\rho(t)||\tau_{\beta})-\frac{1}{\beta}\ln\mathcal{Z}$. Using this definition, \eqref{3} can be generalized for R\'{e}yni divergence as: $\beta\frac{d}{dt}\mathcal{F}^{\gamma}(t)=-\sigma^{\gamma}(t).$ This result shows that negative EPR is necessary and sufficient for free energy backflow.

\textbf{Corollary 2:} \textit{Under thermal operations $\Lambda_{th}$, which is CP-divisible, EPR is always positive. }

\proof: To prove the above corollary, we use the following two facts.
\begin{enumerate}
\item Monotonicity of relative entropy under CPTP maps: $S(\Lambda(\rho)||\Lambda(\sigma))\leq S(\rho||\sigma)$.
\item Thermal state $\tau_{\beta}$ is a fixed point under thermal operation: $\Lambda_{th}(\tau_{\beta})=\tau_{\beta}$.
\end{enumerate}
We have 
\[
\begin{array}{ll}
S(\rho(t)||\tau_{\beta})\geq S(\Lambda_{th}(t+\delta,t)(\rho(t))||\Lambda_{th}(t+\delta)(\tau_{\beta}))\\
=S(\Lambda_{th}(t+\delta,t)(\rho(t))||\tau_{\beta})
\end{array}
\]
Therefore, we have
\[
\begin{array}{ll}
\frac{d}{dt}S(\rho(t)||\tau_{\beta})=\lim_{\delta\rightarrow 0}\frac{S(\rho(t+\delta)||\tau_{\beta})-S(\rho(t)||\tau_{\beta})}{\delta}\leq 0\\
\end{array}
\]
which shows $\sigma(t)=-\frac{d}{dt}S(\rho(t)||\tau_{\beta})\geq 0$, under thermal operations. \qed 

When CP-divisibility breaks down, EPR can be negative and consequently the free energy of the system increases. Evidently, NM acts as a resource and provides free energy to the system. The free energy is a monotone under thermal operations, which means that the system monotonically goes towards the thermal state. We see that NM backflow essentially drives the system away from equilibrium. Therefore, \textbf{Corollary 1} is also true for thermal operation. 

\section{Beyond thermal operations and formulation of GEPR}

The entropy production rate is however, not a positive quantity for all divisible operations which are not thermal. This follows from the fact that for operations which are not thermal, the state $\tau_{\beta}$ is not a fixed point any more, and hence, we have $S(\Lambda(\rho(t) )||\Lambda(\tau_{\beta}))\neq S(\Lambda(\rho(t))||\tau_{\beta})$. In order to investigate such
situations, let us first define the notion of  athermality. \\
\textbf{Athermality:} \textit{We define athermality between the instantaneous state $\rho(t)$ and the thermal state as 
$
\mathcal{A}(t)=D_T(\rho(t),\tau_{\beta}),
$
where $D_T(A,B)=\frac{1}{2}Tr|A-B|$ is the trace distance between two states $A$ and $B$.} \\
Since $\mathcal{A}(t)$ is a monotone under divisible thermal operation, it is also a witness for NM backflow. In the following Theorem, we  establish a complementary relation between NM and free energy loss.\\

\noindent\textbf{Theorem 3:} \textit{Loss of free energy and athermality obeys the following complementary relation:
\beq\label{8}
\Delta\mathcal{F}^{\gamma}(t)+2\gamma\mathcal{A}^2(t) \leq S^{\gamma}(\rho(0)||\tau_{\beta}),~~\forall \gamma \in (0,1]
\eeq
}
\proof: Loss of generalized free energy can be defined as 
$\Delta\mathcal{F}^{\gamma}(t)=\left[S^{\gamma}(\rho(0)||\tau_{\beta})-S^{\gamma}(\rho(t)||\tau_{\beta})\right].$ From the Pinsker inequality \citep{pinsker1} for generalised Reyni divergence \citep{pinsker2} : 
$S^{\gamma}(A||B)\geq 2\gamma D_T(A,B)^2~~\forall \gamma \in (0,1]$,
we get the relation \eqref{8}. The relation naturally also holds for von-Neumann relative entropy ($\gamma\rightarrow 1$).\qed 

 Therefore it is evident that the loss of free energy $\Delta\mathcal{F}^{\gamma}(t)$ can only decrease when the athermality of the system increases, which can only happen under NM backflow of information. This shows that the complementary relation \eqref{8} further bolsters the importance of NM as a resource in various quantum thermodynamic protocols.

We now consider operations $\Lambda_G$, which are beyond thermal operation and generally do not possess any definite long time limit. For such operations the thermal state $\tau_{\beta}$ is not a fixed point any more, since the backaction of bath can produce a time-dependent shift in the Hamiltonian, or external driving Hamiltonians may also be present. Consider a general time-dependent shift in the Hamiltonian,  $H\rightarrow \tilde{H}(t)$ under the evolution $\Lambda_{G}$. Consequently,  the thermal state is modified to a time-dependent thermal state $\tau_{\beta}(t)=\frac{e^{-\beta\tilde{H}(t)}}{\mathcal{Z}(t)}$. We then have $\Lambda_G(t+\delta,t)(\tau_{\beta}(t))=\tau_{\beta}(t+\delta)$. We define a generalized entropy production rate (GEPR) for such evolutions as 
\beq\label{9}
\tilde{\sigma}(t)=-\frac{d}{dt}S(\rho(t)||\tau_{\beta}(t)).
\eeq 
Thermodynamics of open quantum systems having no definite long time limit, has been considered in several earlier works \citep{latest4,latest5,latest6}. The generalization of entropy production rate that we provide here in this work is explicitly  constructed to deal with such non-equilibrium situations.
This GEPR is related to EPR by  
\beq\label{10}
\tilde{\sigma}(t)=\sigma(t)-\beta(\bra W\ket-\bra W\ket_{th}),
\eeq
where $\bra W\ket=Tr[\dot{\tilde{H}}(t)\rho(t)]$ and $\bra W\ket_{th}=Tr[\dot{\tilde{H}}(t)\tau_{\beta}(t)]$ are the workdone by $\rho(t)$ and $\tau_{\beta}(t)$ respectively. 
The proof of \eqref{10} is as follows.
We have 
$
\tilde{\sigma}(t)=-\frac{d}{dt}\left[\rho(t)\ln\rho(t)-\rho(t)\ln\tau_{\beta}(t)\right]
=\frac{d}{dt}S(t)-\beta\mathcal{J}(t)-\beta Tr[\rho(t)\dot{\tilde{H}}(t)]-\frac{d}{dt}\ln\mathcal{Z}(t)
=\sigma(t)-\beta\left(\bra W\ket-\bra W\ket_{th}\right).
$

Further, on the lines of Eq. \eqref{2}, we derive a similar expression for GEPR, given by
\beq\label{10a}
\tilde{\sigma}(t)=\frac{d}{dt}\left[ (S(t)-S_{th}(t))-\beta\mathcal{W}(t)\right],
\eeq
where $S_{th}(t)$ is the von-Neumann entropy of $\tau_{\beta}(t)$. The proof of \eqref{10a} is as follows.
\[
\begin{array}{ll}
S(\rho(t)||\tau_{\beta}(t))=Tr[\rho(t)\ln\rho(t)]-Tr[\rho(t)\ln\tau_{\beta}(t)],\\
=-(S(t)-S_{th}(t))+\beta Tr[\tilde{H}(t)(\rho(t)-\tau_{\beta}(t))].
\end{array}
\]
Therefore, differentiating the above equation with respect to time, we get
$\tilde{\sigma}(t)=\frac{d}{dt}\left[ (S(t)-S_{th}(t))-\beta\mathcal{W}(t)\right]$.
 Based on these findings, we now prove the following corollary for GEPR.

\textbf{Corollary 3:} GEPR is negatively proportional to the time rate of change of the difference between the free energies of the state $\rho(t)$ and $\tau_{\beta}(t)$: %\hspace{0.2cm}  
\hspace{0.2cm}
$
\beta\frac{d}{dt}\left(\mathcal{F}(t)-\mathcal{F}_{th}(t)\right)=-\tilde{\sigma}(t).
$\\
\proof: Differentiating the free energy $\mathcal{F}(t)$ with respect to time, we find 
\[\beta\left(\frac{d}{dt}\mathcal{F}(t)-\bra W\ket_{th}\right)=-\tilde{\sigma}(t).\]
The free energy of the instantaneous thermal state is $\mathcal{F}{\color{red}_{th}}(\rho_{th}(t))=-\frac{1}{\beta}\ln\mathcal{Z}(t)$. By differentiating with respect to time we get $\frac{d}{dt}\mathcal{F}(\rho_{th}(t))=\bra W\ket_{th}$. Hence, the modified relation between the free energy rate and the GEPR is given by
\[
\beta\frac{d}{dt}\left(\mathcal{F}(t)-\mathcal{F}_{th}(t)\right)=-\tilde{\sigma}(t),
\]
where $\mathcal{F}_{th}(t)=-\frac{1}{\beta}\ln\mathcal{Z}(t)$ is the free energy of the instantaneous thermal state $\tau_{\beta}(t)$. \qed

It follows that the negativity of GEPR implies that the system is free energetically going away from the instantaneous thermal state.  
Similar to the case of thermal operation, we establish the following complementary relation.\\

\noindent\textbf{Corollary 4:} \textit{A complementary relation of the form:
$
\Delta\bar{\mathcal{F}} +2\mathcal{A}^2(t) \leq S(\rho(0)||\tau_{\beta}(0)),
$
exists for operations $\Lambda_G$, where $\bar{\mathcal{F}}(t)=(\mathcal{F}(\rho(t))-\mathcal{F}(\tau_{\beta}(t)))$ is the free energy difference between the state $\rho(t)$ and $\tau_{\beta}(t)$, $\mathcal{A}(t)=D_T(\rho(t)||\rho_{th}(t))$ is the instantaneous athermality and $\Delta\bar{\mathcal{F}}(t)=(\bar{\mathcal{F}}(0)-\bar{\mathcal{F}}(t))$.}

\noindent The proof is similar to that of Theorem 2.

The consequences of \textbf{Corollary 3} and \textbf{4} are rather similar to that of what we found for thermal operations. They show that NM backflow is necessary to drive the system away from its instantaneous equilibrium state and hence, is indispensable for regenerating the resource, which is the free energetic difference between the state and the thermal state.  In the following section, we consider a realistic example for a central spin system, to validate our theory of GEPR.

%%%%%%%%%%%%%%%%%%%%%%%%%%%%%%%%%%%%%%%%%%%%%%%%%%%%%%%%%%%%%%%%%%%%%%%%%%%%%%%%%%%%%%%%%%%%%%
%%%%%%%%%%%%%%%%%%%%%%%%%%%  up to here %%%%%%%%%%%%%%%%%%%%%%%%%%%%%%%%%%%%%%%%%
%%%%%%%%%%%%%%%%%%%%%%%%%%%%%%%%%%%%%%%%%%%%%%%%%%%%%%%%%%%%%%%%%%%%%%%%%%%%%%%%%%%%%%%%%

\section{Example of a spin bath model}

 Here we examine the validity of our findings for the $\Lambda_G$ operation in the backdrop of a spin-bath model. The model consists of a single spin interacting with $N$ number of mutually non-interacting spin-half particles. The collection of non-interacting spins is considered to be the bath. This type of fermionic bath model has been of significant interest for over the past decade \citep{spin1,spin2} and extremely relevant for quantum computing
with NV centre \citep{spin3} defects within a diamond lattice.

%Details of the mentioned quantities and corresponding calculations are given in the supplimentary material \citep{sup}.

Let the Hamiltonian corresponding to the system, bath and their interaction be given by $\tilde{H}_S, \tilde{H}_B, \tilde{H}_I$ respectively. The total Hamiltonian $\tilde{H}$ is given by,

%The total Hamiltonian of the system, environment and the system-environment interaction is represented as
\beq\label{sec2a}
\tilde{H}=\tilde{H}_S+\tilde{H}_B+\tilde{H}_I,
\eeq
where the system, environment and interaction Hamiltonians $\tilde{H}_I$ are respectively given by
\beq\label{sec2b}
\begin{array}{ll}
\tilde{H}_S=\hbar g\omega_0\sigma_z,\\
\tilde{H}_B=\hbar g\frac{\omega}{N}\sum_{i=1}^N\sigma_z^i,\\
\tilde{H}_I=\hbar g\frac{\alpha}{\sqrt{N}}\sum_{i=1}^N \left(\sigma_x\sigma_x^i+\sigma_y\sigma_y^i+\sigma_z\sigma_z^i\right),
\end{array}
\eeq
where $\sigma_{k}$, $k=x,y,z$ are the Pauli matrices, with the superscript `i' stands for the i-th particle of the bath. $g$ is a constant with the dimension of frequency, $\omega_{0}$ and $\omega$ are the dimensionless parameters respectively characterizing the difference of energy levels
of the system and the environment. $\alpha$ is the system-environment coupling strength.
Utilizing the total angular momentum of the bath spin particles $J_{k}=\sum\limits_{i=1}^N\sigma^{i}_{k}$, and  using the Holstein-Primakoff transformation, given by
\[  
J_+=\sqrt{N}b^{\dagger}\left(1-\frac{b^{\dagger}b}{2N}\right)^{1/2}~~,~~J_-=\sqrt{N}\left(1-\frac{b^{\dagger}b}{2N}\right)^{1/2}b,
\]
the bath and the system-environment interaction Hamiltonians can be rewritten as 
\beq\label{sec2c}
\begin{array}{ll}
\tilde{H}_B~~=-\hbar g\omega\left(1-\frac{b^{\dagger}b}{N}\right),\\
\tilde{H}_{I}=2\hbar g\alpha\left[\sigma_{+}\left(1-\frac{b^{\dagger}b}{2N}\right)^{1/2}b+\sigma_{-}b^{\dagger}\left(1-\frac{b^{\dagger}b}{2N}\right)^{1/2}\right]\\
~~~~~-\hbar g\alpha\sqrt{N}\sigma_z\left(1-\frac{b^{\dagger}b}{N}\right).
\end{array}
\eeq
Here $b$ and $b^{\dagger}$ are bosonic annihilation and creation operators respectively. We take the initial system-bath state as $\rho_S(0)\otimes\rho_B(0)$.  The initial system qubit is considered as $\rho_S(0)=\rho_{11}(0)|1\ket\bra 1|+\rho_{22}(0)|0\ket\bra 0|+\rho_{12}(0)|1\ket\bra 0|+\rho_{21}(0)|0\ket\bra 1|$, whereas the initial environment state is taken to be a thermal state $\rho_B(0)=\exp(-\tilde{H}_B/K\tilde{T})$ with an arbitrary temperature $\tilde{T}$, where K is the Boltzmann constant. The reduced dynamics of the system state can then be calculated \citep{samya3} as
$\rho_S(t)=\mbox{Tr}_B\left[\exp\left(-iHt\right)\rho_S(0)\otimes\rho_B(0)\exp\left(iHt\right)\right]$. Here
\[H=\frac{\tilde{H}}{\hbar g},~~t=g\tilde{t},~~\mbox{and}~~T=\frac{K\tilde{T}}{\hbar g},\]
where $H$, $t$ and $T$ are all dimensionless quantities. Solving the global Schr\"{o}dinger equation corresponding to the above mentioned Hamiltonian, we get
\beq\label{A1}
\begin{array}{ll}
\rho_{11}(t)=\rho_{11}(0)(1-A(t))+\rho_{22}(0)B(t),\\
\rho_{12}(t)=\rho_{12}(0)C(t),
\end{array}
\eeq
with
\[
\begin{array}{ll}
A(t)=\sum_{n=0}^N (n+1)\alpha^2(1-n/2N)\left(\frac{\sin(\eta t/2)}{\eta/2}\right)^2\frac{e^{-\frac{\omega}{T}(n/N-1)}}{Z},\\
\\
B(t)=\sum_{n=0}^N n\alpha^2(1-(n-1)/2N)\left(\frac{\sin(\eta' t/2)}{\eta'/2}\right)^2\frac{e^{-\frac{\omega}{T}(n/N-1)}}{Z},\\
\\
C(t)=\sum_{n=0}^N e^{-i(\Lambda -\Lambda')t/2}\left(\cos(\eta t/2)-i\frac{\theta}{\eta}\sin (\eta t/2)\right)\\
~~~~~~~~\times \left(\cos(\eta' t/2)+i\frac{\theta'}{\eta'}\sin (\eta' t/2)\right)\frac{e^{-\frac{\omega}{T}(n/N-1)}}{Z},\\
\\
Z=\sum_{n=0}^Ne^{-\frac{\omega}{T}(n/N-1)},\\
\\
\eta=2\sqrt{\left(\omega_0-\frac{\omega}{2N}-\alpha\sqrt{N}\left(1-\frac{2n+1}{2N}\right)\right)^2+4\alpha^2(n+1)(1-\frac{n}{2N})},\\
\\
\eta'=2\sqrt{\left(\omega_0-\frac{\omega}{2N}-\alpha\sqrt{N}\left(1-\frac{2n-1}{2N}\right)\right)^2+4\alpha^2 n(1-\frac{(n-1)}{2N})},\\
\\
\theta=2\left(\omega_0-\omega/2N+\alpha\sqrt{N}\left(1-\frac{2n+1}{2N}\right)\right),\\
\\
\end{array}
\]

\[
\begin{array}{ll}
\theta'=-2\left(\omega_0-\omega/2N-\alpha\sqrt{N}\left(1-\frac{2n-1}{2N}\right)\right),\\
\Lambda=-2\omega\left(1-\frac{2n+1}{2N}\right)-\frac{\alpha}{\sqrt{N}},\\
\Lambda'=-2\omega\left(1-\frac{2n-1}{2N}\right)-\frac{\alpha}{\sqrt{N}}.
\end{array}
\]

The master equation for the reduced dynamics presented above \citep{sam2}, is given by
$
\dot{\rho}(t)=\frac{i}{\hbar}U(t)[\rho(t),\sigma_z]+\Gamma_{deph}(t)[\sigma_z\rho(t)\sigma_z-\rho(t)]+\Gamma_{dis}(t)[\sigma_{-}\rho(t)\sigma_{+}-\frac{1}{2}\{\sigma_{+}\sigma_{-},\rho(t)\}]+\Gamma_{abs}(t)[\sigma_{+}\rho(t)\sigma_{-}-\frac{1}{2}\{\sigma_{-}\sigma_{+},\rho(t)\}]
$,
where $\sigma_{\pm}=\frac{\sigma_x \pm i\sigma_y}{2}$, and $\Gamma_{dis}(t), \Gamma_{abs}(t), \Gamma_{deph}(t)$ are the rates of dissipation, absorption and dephasing processes respectively, and  $U(t)$ corresponds to the unitary evolution.\\

\textbf{The Lindblad coefficients:} The rates of dissipation, absorption, dephasing and the unitary evolution are, respectively, given as
\beq\label{sec1I}
\begin{array}{ll}
\Gamma_{dis}(t)=\left[\frac{d}{dt}\frac{(A(t)-B(t))}{2}-\frac{(A(t)-B(t)+1)}{2}\frac{d}{dt}\ln(1-A(t)-B(t))\right],\\
\\
\Gamma_{abs}(t)=-\left[\frac{d}{dt}\frac{(A(t)-B(t))}{2}-\frac{(A(t)-B(t)-1)}{2}\frac{d}{dt}\ln(1-\alpha(t)-\beta(t))\right],\\
\\
\Gamma_{deph}(t)=\frac{1}{4}\frac{d}{dt}\left[\ln\left(\frac{1-A(t)-B(t)}{|C(t)|^2}\right)\right],\\
\\
U(t)=-\frac{1}{2}\frac{d}{dt}\left[\ln\left(1+\left(\frac{C_R(t)}{C_I(t)}\right)^2\right)\right].
\end{array}
\eeq

The system Hamiltonian evolves to the time dependent $\tilde{H}(t)=U(t)\sigma_Z$, due to the back action of the bath. 
\begin{figure}[htb]
	{\centerline{\includegraphics[width=6cm, height=5cm] {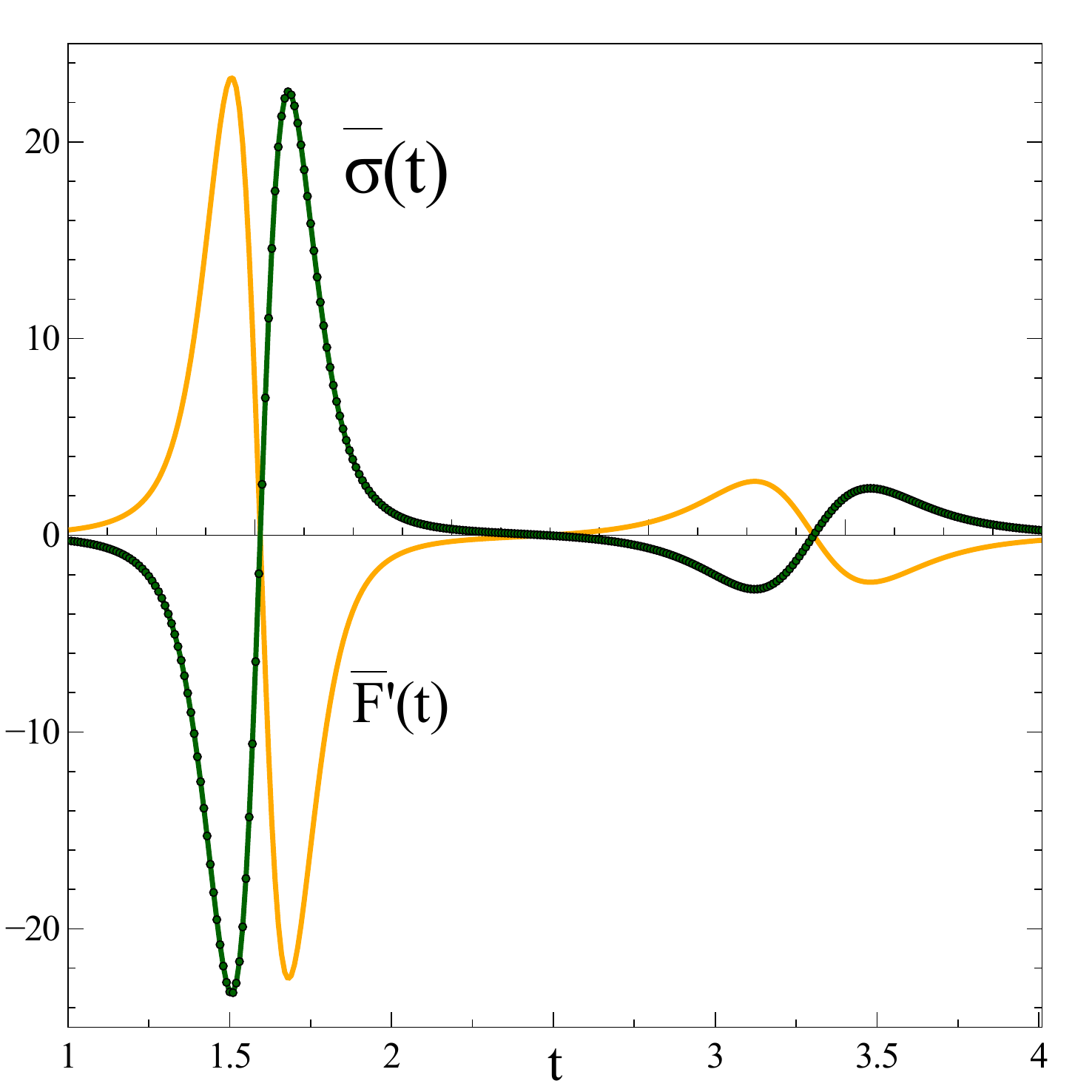}}}
	\caption{(Colour online)\\ Here we plot $\bar{F}'(t)=\frac{d}{dt}\left(\mathcal{F}(t)-\mathcal{F}(\tau_{\beta}(t))\right)$ and $\tilde{\sigma}(t)$ vs. time $t$ with setting the parameters $\beta=1$, $\omega_0=\omega=1$ and $\alpha=0.1$. All quantities are dimensionless. The plot shows that the modified relation between GEPR and rate of free energy change given in \textbf{Corollary 3} is accurate.  }
	\label{fig1}
\end{figure}
\begin{figure}[htb]
	{\centerline{\includegraphics[width=6cm, height=5cm] {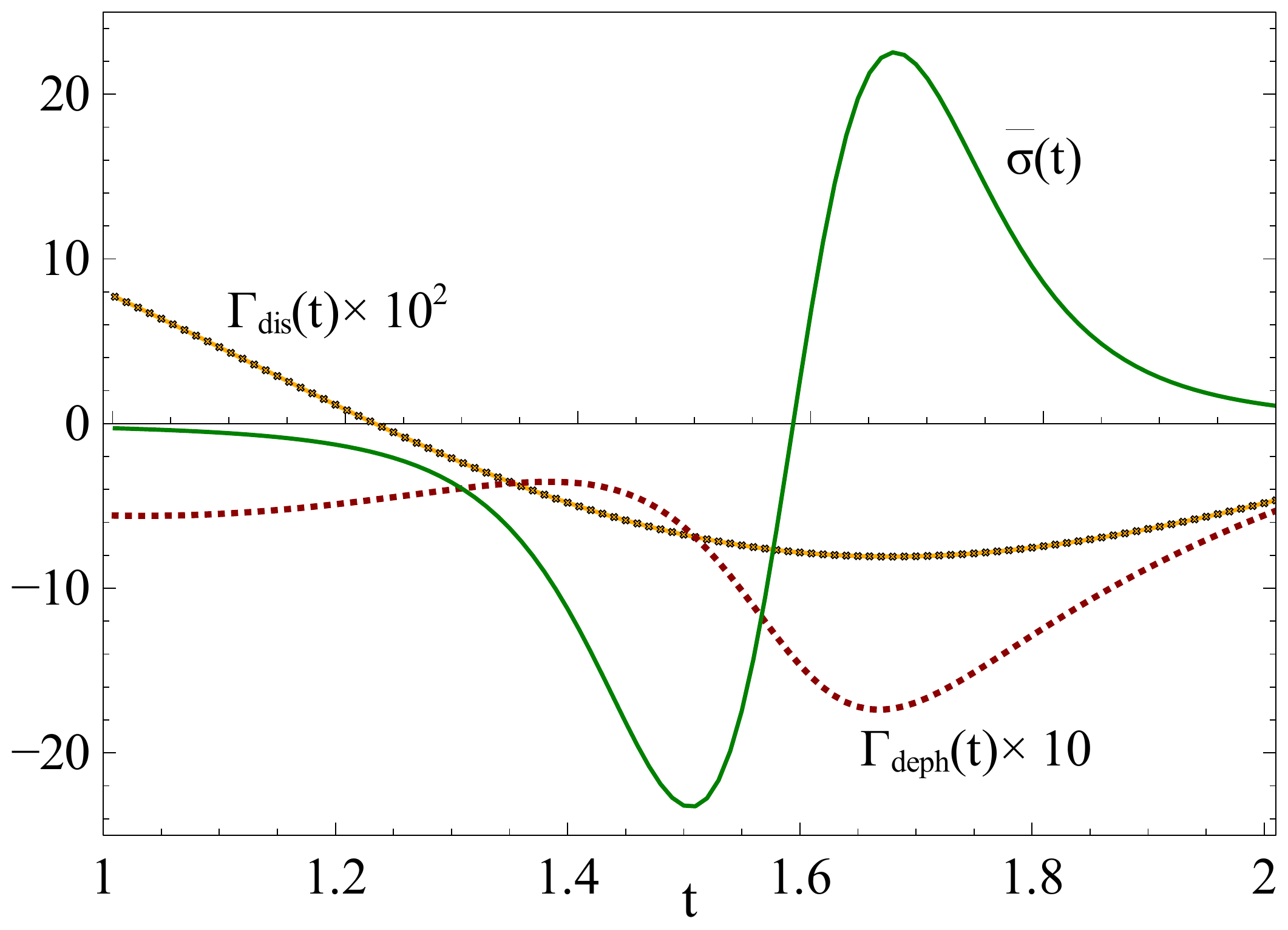}}}
	\caption{(Colour online) \\ Here we plot $\tilde{\sigma}(t)$,  $
	\Gamma_{dis}(t)$ and $\Gamma_{deph}(t)$ vs time $t$ with setting the parameters $\beta=1$, $\omega_0=\omega=1$ and $\alpha=0.1$. All quantities are dimensionless.}
	\label{fig2}
\end{figure}
In Fig.\eqref{fig1} we show that the relation between GEPR and the rate of change of free energy given in \textbf{Corollary 3} holds perfectly. In Fig.\eqref{fig2} we see that GEPR can only be negative when CP-divisibility breaks down ($\Gamma_{dis}(t)<0~\mbox{and (or)}~\Gamma_{deph}(t)<0$). But we also see that in some NM region, GEPR is positive, proving NM is necessary (though may not sufficient) to drive the system away from equilibrium.

\section{Conclusions}

In this paper we have investigated thermodynamic advantages of non-Markovianity from the  perspective of resource interconversion. Considering resource theories of purity and thermodynamics, as well
as a more general non-equilibrium scenario as frameworks, we have shown that non-Markovianity is
related to the concept of the entropy production rate. In case of the resource theory of purity,
we show that non-Markovianity is necessary to drive the system away from the equilibrium. We have
further defined athermality  in the resource theory of thermodynamics, and showed that the loss of free energy  and athermality obey a complementary relationship. Using this fact we establish that the free energy can increase only under non-Markovian backflow of information. Similar results are 
shown to hold even if we go beyond thermal operations. 

An important feature of our analysis is the classification of Lindblad generators for specific quantum operations like unital and thermal operations.
We have been able to fully characterize the Lindblad generators corresponding to unital operations. In case of thermal operations, characterization of Lindblad generators has been done for some special cases which includes the elementary thermal operations. Importantly, they encompasses such thermal operations that are experimentally implementable, for example, thermal operations which can be modelled by the Jaynes-Cummings interactions. 

We have shown for all the considered cases that the backflow of resource happens when EPR (GEPR) is negative due to non-Markovianity. We have given specific examples, including one from a spin-bath model in order to validate our findings. Interpreting non-Markovian memory effects in connection with revivals of purity or the free energy unveils a linkage between open quantum systems and thermodynamics. 
This work entailing the characterization of various aspects of non-Markovian dynamics in terms of inter-convertibility of different quantum resources thus takes an essential step towards exploring 
further connections between the theory of open quantum systems and quantum thermodynamics.  

\bibliographystyle{apsrev4-1}
\bibliography{sister1}

\end{document}